\documentstyle[axodraw,12pt]{article}

\catcode`@=11
\newcount\@tempcntc
\def\@citex[#1]#2{\if@filesw\immediate\write\@auxout{\string\citation{#2}}\fi
  \@tempcnta\z@\@tempcntb\m@ne\def\@citea{}\@cite{\@for\@citeb:=#2\do
    {\@ifundefined
       {b@\@citeb}{\@citeo\@tempcntb\m@ne\@citea\def\@citea{,}{\bf ?}\@warning
       {Citation `\@citeb' on page \thepage \space undefined}}%
    {\setbox\z@\hbox{\global\@tempcntc0\csname b@\@citeb\endcsname\relax}%
     \ifnum\@tempcntc=\z@ \@citeo\@tempcntb\m@ne
       \@citea\def\@citea{,}\hbox{\csname b@\@citeb\endcsname}%
     \else
      \advance\@tempcntb\@ne
      \ifnum\@tempcntb=\@tempcntc
      \else\advance\@tempcntb\m@ne\@citeo
      \@tempcnta\@tempcntc\@tempcntb\@tempcntc\fi\fi}}\@citeo}{#1}}
\def\@citeo{\ifnum\@tempcnta>\@tempcntb\else\@citea\def\@citea{,}%
  \ifnum\@tempcnta=\@tempcntb\the\@tempcnta\else
   {\advance\@tempcnta\@ne\ifnum\@tempcnta=\@tempcntb \else \def\@citea{--}\fi
    \advance\@tempcnta\m@ne\the\@tempcnta\@citea\the\@tempcntb}\fi\fi}
\catcode`@=12
\begin{document}

\title{\vskip-3cm{\baselineskip14pt
\centerline{\normalsize DESY 99-099\hfill ISSN 0418-9833}
\centerline{\normalsize MPI/PhT/99-029\hfill}
\centerline{\normalsize hep-ph/9907489\hfill}
\centerline{\normalsize July 1999\hfill}}
\vskip1.5cm
Ultrasoft Effects in Heavy-Quarkonium Physics}
\author{{\sc Bernd A. Kniehl}\thanks{Permanent address: II. Institut f\"ur 
Theoretische Physik, Universit\"at Hamburg, Luruper Chaussee 149,
22761 Hamburg, Germany.}\\
{\normalsize Max-Planck-Institut f\"ur Physik (Werner-Heisenberg-Institut),}\\
{\normalsize F\"ohringer Ring 6, 80805 Munich, Germany}\\
\\
{\sc Alexander A. Penin}\thanks{Permanent address:
Institute for Nuclear Research, Russian Academy of Sciences,
60th October Anniversary Prospect 7a, Moscow 117312, Russia.}\\
{\normalsize II. Institut f\"ur Theoretische Physik, Universit\"at Hamburg,}\\
{\normalsize Luruper Chaussee 149, 22761 Hamburg, Germany}}

\date{}

\maketitle

\thispagestyle{empty}

\begin{abstract}
In the framework of nonrelativistic QCD, we consider a new class of radiative
corrections, which are generated at next-to-next-to-next-to-leading order
through the chromoelectric dipole interaction of heavy quarkonium with
ultrasoft virtual gluons.
We provide analytical formulae from which the resulting shifts in the
quarkonium energy levels and the wave functions at the origin may be
calculated.
We discuss the phenomenological implications for the top-antitop and 
$\Upsilon$ systems and point out some limitations of describing charmonium 
using a Coulomb potential.

\medskip

\noindent
PACS numbers: 12.38.Bx, 12.39.Jh, 14.40.Gx
\end{abstract}

\newpage

\section{Introduction}

Recently, essential progress has been made in the theoretical investigation
of the pair production of heavy quarks at threshold, and analytical results
are now available up to next-to-next-to-leading order (NNLO) 
\cite{KPP,MelYel,PP,Hoa,HoaTeu,BSS,Nag,BenSin}.
The NNLO corrections have turned out to be so sizeable that it appears to be
indispensible to also gain control over the next-to-next-to-next-to-leading
order (N$^3$LO), both in regard of phenomenological applications and in order 
to understand the structure and the peculiarities of the threshold expansion.
In this paper, we take the first step in this direction and investigate a 
particular class of N$^3$LO corrections, namely those arising from the
emission and absorbtion of virtual ultrasoft gluons by the heavy quarks.
Such ultrasoft corrections are absent in lower orders and represent a
genuinely new feature of the N$^3$LO.

Specifically, we study the near-threshold behavior of the vacuum-polarization
function $\Pi(q^2)$ of a heavy-quark vector current $j_\mu=\bar q\gamma_\mu q$,
\begin{equation}
\left(q_\mu q_\nu-g_{\mu \nu}q^2\right)\Pi(q^2)
=i\int d^4xe^{iq\cdot x}\langle 0|Tj_{\mu}(x)j_{\nu}(0)|0\rangle\,.
\end{equation}
Its imaginary part gives (up to a constant factor) the spectral density of
$q\bar q$ production in $e^+e^-$ annihilation. 
Near threshold, the heavy quarks are nonrelativistic, so that one may consider
the quark velocity $\beta$ (or inverse quark mass) as a small parameter.
An expansion in $\beta$ may be performed directly in the Lagrangian of quantum
chromodynamics (QCD) by using the framework of effective field theory.
In the threshold problem, there are four different scales \cite{BenSmi}:
(i) the hard scale (energy and momentum scale like $m_q$);
(ii) the soft scale (energy and momentum scale like $\beta m_q$);
(iii) the potential scale (energy scales like $\beta^2 m_q$, while 
momentum scales like $\beta m_q$); and
(iv) the ultrasoft scale (energy and momentum scale like $\beta^2 m_q$).
The ultrasoft scale is only relevant for gluons.
  
By integrating out the hard scale of QCD, one arrives at the effective theory
of nonrelativistic QCD (NRQCD) \cite{CasLep}. 
If one also integrates out the soft scale and the potential gluons, one 
obtains the effective theory of potential NRQCD (pNRQCD), which contains
potential quarks and ultrasoft gluons as active particles \cite{PinSot1}.
The dynamics of the quarks is governed by the effective, nonrelativistic 
Schr{\"o}dinger equation and by their interaction with the ultrasoft gluons.
To get a regular perturbative expansion within pNRQCD, this interaction should
be expanded in multipoles.
The corrections from harder scales are contained in the Wilson coefficients,
leading to an expansion in $\alpha_s$, as well as in the higher-dimensional
operators of the nonrelativistic Hamiltonian, corresponding to an expansion in
$1/m_q$.

This paper is organized as follows.
In Section~2, we introduce the basic features of the pNRQCD formalism and
derive formulas from which the leading ultrasoft corrections to the energy
levels and the wave functions at the origin of heavy quarkonia may be
evaluated.
In Section~3, we present a numerical analysis and discuss phenomenological
implications of our results.
Section~4 contains our conclusions.
In the Appendix, we explain how the ``QCD Bethe logarithms'' introduced in 
Section~2 may be reduced to one-dimensional integrals of elementary 
functions.

\section{Ultrasoft corrections to energy levels and wave functions}

In this section, we briefly recall the formalism of pNRQCD and calculate the
leading ultrasoft corrections to the energy levels and the wave functions at
the origin of heavy quarkonia.
The basic quantity of pNRQCD is the nonrelativistic Green function, $G^{s,o}$,
of the Schr{\"o}dinger equation,
\begin{equation}
\left({\cal H}^{s,o}-E\right)G^{s,o}({\bf x},{\bf y},E)
=\delta^{(3)}({\bf x}-{\bf y})\,,
\end{equation}
where ${\cal H}^{s,o}$ is the nonrelativistic Hamiltonian for the quark pair
in the colour-singlet (s) [colour-octet (o)] state, defined by
\begin{eqnarray}
{\cal H}^{s,o}&=&-{{\bf \Delta}_{\bf x}\over m_q}+V^{s,o}(x)+\ldots,
\nonumber\\
V^{s,o}(x)&=&V^{s,o}_C(x)+\ldots,
\end{eqnarray}
with ${\bf \Delta_x}=\partial_{\bf x}^2$ and $x=|{\bf x}|$.
The ellipses stand for the higher-order terms in $\alpha_s$ and $1/m_q$.
The Coulomb (C) potentials for the singlet and octet states are attractive and
repulsive, respectively, and are given by
\begin{eqnarray}
V_C^s(x)&=&-C_F\frac{\alpha_s}{x},
\nonumber\\
V_C^o(x)&=&\left(\frac{C_A}{2}-C_F\right)\frac{\alpha_s}{x},
\end{eqnarray}
where $C_A=3$ and $C_F=4/3$ are the eigenvalues of the quadratic Casimir
operators of the adjoint and fundamental representations of the colour group,
respectively.
 
The propagation of the quark-antiquark pair in the singlet and octet states is
described by the singlet and octet Green functions, respectively, which have
the following spectral representations \cite{VolLeu}:
\begin{eqnarray}
G^s({\bf x},{\bf y},E)&=&\sum_{n=1}^\infty{\psi^*_n({\bf x})\psi_n({\bf y})
\over E_n-E}+
\int_0^\infty{d^3k\over(2\pi)^3}\,
{\psi^{s*}_{\bf k}({\bf x})\psi^s_{\bf k}({\bf y})\over k^2/m_q-E}\,,
\nonumber\\
G^o({\bf x},{\bf y},E)&=&\int_0^\infty{d^3k\over(2\pi)^3}
{\psi^{o*}_{\bf k}({\bf x})\psi^o_{\bf k}({\bf y})\over k^2/m_q-E}\,,
\end{eqnarray}
where $\psi_m$ and $\psi_{\bf k}^{s,o}$ are the wave functions of the
$q\bar q$ bound and continuum states.
Note that a discrete part of the spectrum (bound states) only exists for the
singlet Green function. 

The nonrelativistic expansion in $\alpha_s$ and $\beta=\sqrt{1-4m_q^2/s}$,
where $\sqrt s$ is the $q\bar q$ centre-of-mass energy, provides us with the
following representation of the heavy-quark vacuum-polarization function near
threshold:
\begin{equation}
\Pi(E)={N_c\over 2 m_q^2}G^s_C(0,0,E)+\ldots,
\end{equation}
where $E=\sqrt{s}-2m_q$ is the $q\bar q$ energy counted from the threshold and
$G^s_C$ is the leading-order Coulomb Green function, which sums up the
$(\alpha_s/\beta)^n$ terms singular near the threshold.
The ellipsis stands for the higher-order terms in $\alpha_s$ and $\beta$.

We are interested in the correction to the Green function induced by a 
virtual ultrasoft gluon.
The corresponding diagram is shown in Fig.~1.
The leading contribution is the one due to the chromoelectric dipole
interaction $g_s({\bf r_q}-{\bf r_{\bar q}})\cdot{\bf E}$ of the
quark-antiquark pair with the ultrasoft gluon \cite{VolLeu}. 
We wish to calculate the corresponding corrections to the energy levels and
the wave functions at the origin of several low-lying resonances, which
represent key parameters for the analysis of the threshold production of
top- and bottom-quark pairs.
Near the $n$th pole of the discrete spectrum, the correction to $G^s_C$ reads 
\begin{equation}
\Delta G(0,0,E)|_{E\rightarrow E_n}=-{|\psi_n(0)|^2\over (E_n-E)^2}J(E)\,,
\end{equation}
where
\begin{equation}
J(E)=C_Fg_s^2\int_0^\infty{d^3k\over(2\pi)^3}
\langle r_i\rangle_{{\bf k}n}\langle r_j\rangle_{n{\bf k}}I^{ij}(E-k^2/m_q)\,,
\label{eq:je}
\end{equation}
with
\begin{equation}
I^{ij}(p)=-i\int{d^Dl\over (2\pi)^D}\,{l_0^2(\delta^{ij}-l^il^j/{\bf l^2})
\over l^2(p-l^0)}\,,
\end{equation}
in $D$ space-time dimensions.
After integration over $l_0$, we recover the well-known nonrelativistic
perturbation theory expression. 
The pole in the nonrelativistic propagator is bypassed according to the
standard prescription $E\rightarrow E+i\varepsilon$. 
The remaining integral over ${\bf l}$ is ultraviolet divergent for $D=4$.
To obtain a finite result, we use dimensional regularization with
$D=4-\epsilon$.
The nonrelativistic perturbation theory of quantum electrodynamics (QED) in
dimensional regularization is comprehensively described in
Refs.~\cite{PinSot2,CMY}.
We thus obtain
\begin{equation}
I^{ij}(p)=p^3{\delta^{ij}\over6\pi}
\left({1\over\bar\epsilon}+\ln{\mu_f\over-p}+{5\over6}-\ln{2}\right)\,,
\end{equation}
where $1/\bar\epsilon=1/\epsilon+[\ln(4\pi)-\gamma_E]/2$.
Note that this divergence is spurious.
It arises in the process of scale separation due to the use of pNRQCD
perturbation theory at short distances where it is inapplicable. 
In the total N$^3$LO result, the pole in $1/\epsilon$ is cancelled by the 
infrared poles coming from the hard- and soft-scale corrections. 
However, since these corrections are still unknown, we subtract the divergent
part according to the $\overline{\rm MS}$ scheme.
This means that the same scheme must be used for the calculation of the hard-
and soft-scale corrections.
As a consequence, the partial result for the ultrasoft contribution depends on
the auxiliary ``factorization scale'' $\mu_f$, which drops out in the total
result.

The matrix element $\langle{\bf r}\rangle_{{\bf k}n}$ is the one between the
singlet Coulomb wave function of principal quantum number $n$ and the octet
Coulomb wave function of momentum ${\bf k}$.
Since this matrix element is finite, we may use the wave functions of the
Schr{\"o}dinger equation in three dimensions, with $\epsilon=0$.
Writing
\begin{eqnarray}
E_n&=&E^C_n+\Delta E_n\,,
\nonumber\\
|\psi_n(0)|^2&=&\left|\psi^C_n(0)\right|^2\left(1+\Delta\psi_n^2\right)\,,
\end{eqnarray}
where the Coulomb values are
\begin{eqnarray}
E^C_n&=&-{\lambda_s^2\over m_qn^2}\,,
\nonumber\\
|\psi^C_n(0)|^2&=&{\lambda_s^3\over n^3}\,,
\end{eqnarray}
with
$\lambda_s=\alpha_sC_Fm_q/2$, we obtain the leading ultrasoft corrections as
\begin{eqnarray}
\Delta E_n&=&J(E)|_{E=E^C_n}\,,
\nonumber\\
\Delta\psi_n^2(0)&=&\left.{\partial J(E)\over\partial E}\right|_{E=E^C_n}\,.
\end{eqnarray}
Inserting Eq.~(\ref{eq:je}), we thus obtain
\begin{eqnarray}
\Delta E_n&=&
-{2C_F\alpha_s\over3\pi}\int_0^\infty{d^3k\over(2\pi)^3}
|\langle {\bf r}\rangle_{{\bf k}n}|^2\left(E^C_n-k^2/m_q\right)^3
\left(\ln{E^C_1\over E^C_n-k^2/m_q}+\ln{\mu_f\over E^C_1}
+{5\over 6}-\ln{2}\right)\,,
\nonumber\\
\Delta\psi_n^2(0)&=&
-{2C_F\alpha_s\over\pi}\int_0^\infty{d^3k\over(2\pi)^3}
|\langle {\bf r}\rangle_{{\bf k}n}|^2\left(E^C_n-k^2/m_q\right)^2 
\left(\ln{E^C_1\over E^C_n-k^2/m_q}+\ln{\mu_f\over E^C_1}+
{1\over 2}-\ln{2}\right)\,.
\nonumber\\
&&\label{eq:de}
\end{eqnarray}
Except for the terms involving $\ln[E^C_1/(E^C_n-k^2/m_q)]$, we may evaluate
the right-hand sides of Eq.~(\ref{eq:de}) analytically.
In fact, making use of the completeness relation,
\begin{equation}
\int_0^\infty{d^3k\over(2\pi)^3}|\langle{\bf r}\rangle_{{\bf k}n}|^2 
(E^C_n-k^2/m_q)^m
=\langle {\bf r}(E^C_n-{\cal H}^o)^m{\bf r}\rangle_{nn},
\end{equation}
this problem may be reduced to the calculation of the diagonal matrix elements
of appropriate local operators,
\begin{eqnarray}
\left\langle{\bf r}\left(E^C_n-{\cal H}^o\right)^3{\bf r}\right\rangle_{nn}
&=&\left\langle r^2(V^s-V^o)^3
+\frac{4}{m_q}(V^s-V^o)\left(E^C_n-3V^s+V^o\right)
-2{\partial^2V^o\over m_q^2}\right\rangle_{nn}
\nonumber\\
&=&E^C_n\left[{1\over 4}C_A^3+{2\over n}C_A^2C_F+
\left({6\over n}-{1\over n^2}\right)C_AC_F^2+
{4\over n}C_F^3\right]\,,
\nonumber\\
\left\langle{\bf r}\left(E^C_n-{\cal H}^o\right)^2{\bf r}\right\rangle_{nn}
&=&\left\langle r^2(V^s-V^o)^2
+\frac{4}{m_q}\left(E^C_n-2V^s+V^o\right)\right\rangle_{nn}
\nonumber\\
&=&\left({1\over 4}C_A^2C_F+{1\over n^2}C_AC_F^2+{1\over n^2}C_F^3\right)\,.
\end{eqnarray}
The $C_F^3$ contribution is purely Abelian and coincides with the QED result
for the positronium bound state \cite{PinSot2}.
The ``maximal non-Abelian'' contribution proportional to $C_A^3$ in the local
part of the energy shift may be read off from the analysis of the infrared
properties of the non-Abelian static potential \cite{ADM}.
The $C_A^3/\epsilon$ term in the ultrasoft contribution to the energy shift
cancels the corresponding infrared pole in the potential \cite{Bra}.
Note that there is no such term in the wave-function correction.
This may be understood by observing that, in contrast to the case of the 
energy shift, the wave function receives an additional infrared
$C_A^3/\epsilon$ contribution from the hard matching coefficient of the 
heavy-quark vector current \cite{CzaBen}.
This hard contribution must cancel the potential-related divergence in the
correction to the wave function, but it does not affect the energy levels.

The logarithmic contributions in Eq.~(\ref{eq:de}) represent a pure
``retardation effect'' and cannot be reduced to local-operator contributions.
It is convenient to introduce the ``QCD Bethe logarithms'',
\begin{eqnarray}
L^E_n&=&{1\over C_F^2E^C_n}\int_0^\infty{d^3k\over(2\pi)^3}
|\langle{\bf r}\rangle_{{\bf k}n}|^2\left(E^C_n-k^2/m_q\right)^3
\ln{E^C_1\over E^C_n-k^2/m_q}\,,
\nonumber\\
L^\psi_n&=&{1\over C_F^2}\int_0^\infty{d^3k\over(2\pi)^3}
|\langle{\bf r}\rangle_{{\bf k}n}|^2\left(E^C_n-k^2/m_q\right)^2
\ln{E^C_1\over E^C_n-k^2/m_q}\,.
\label{eq:be}
\end{eqnarray}
In the Appendix, we explain how these QCD Bethe logarithms may be reduced to
one-dimensional integrals of elementary functions.
For $n=1,2,3$, we obtain the following numerical values:
\begin{equation}
\begin{array}{lll}
L^E_1=-81.5379\,,\qquad&L^E_2=-37.6710\,,\qquad&L^E_3=-22.4818\,,\\
L^\psi_1=-5.7675\,,\qquad&L^\psi_2=0.7340\,,\qquad&L^\psi_3=2.2326\,.\\
\end{array}
\end{equation}
The final results for the ultrasoft corrections to the heavy-quarkonium energy
levels and wave functions at the origin read
\begin{eqnarray}
\Delta E_n&=&-{2\alpha_s^3\over 3\pi}E^C_n\left\{
\left[{1\over 4}C_A^3+{2\over n}C_A^2C_F+
\left({6\over n}-{1\over n^2}\right)C_AC_F^2+
{4\over n}C_F^3\right]\right.
\nonumber\\
&&{}\times\left.
\left(\ln{\mu_f\over E^C_1}+{5\over6}-\ln{2}\right)+C_F^3L^E_n\right\}\,,
\nonumber\\
\Delta\psi^2_n&=&-{2\alpha_s^3\over\pi}\left[
\left({1\over4}C_A^2C_F+{1\over n^2}C_AC_F^2+
{1\over n^2}C_F^3\right)\left(\ln{\mu_f\over E^C_1}+{1\over 2}-\ln{2}\right)
+C_F^3L^\psi_n\right]\,.
\label{eq:fin}
\end{eqnarray}

\section{Phenomenological applications}

We are now in a position to present a numerical analysis and to discuss the
phenomenological implications of our results.
As illustrative examples, we consider, in Figs.~2 and 3, the $b\bar b$ and
$t\bar t$ ground states, with $n=1$, respectively, and investigate the size
and the dependence on the factorization scale $\mu_f$ of the ultrasoft
corrections to the energy level $E_1$ and the square of the wave function at
the origin $|\psi_1(0)|^2$.
As input values for the pole masses and the strong coupling constant, we use
$m_b=4.8$~GeV, $m_t=175$~GeV and $\alpha_s(M_Z)=0.118$, respectively.
Note that, in Eq.~(\ref{eq:fin}), one power of $\alpha_s$ refers to the
ultrasoft gluon interaction and should be evaluated at the ultrasoft scale,
$\alpha_s^2m_q$, while the two residual powers of $\alpha_s$ originate from
the Coulomb Green function and should be evaluated at the Coulomb scale,
$\alpha_sm_q$.
This leads us to solve the functional equations
$\alpha_s(\alpha_s^2m_q)=\alpha_s$ and $\alpha_s(\alpha_sm_q)=\alpha_s$ for
the ultrasoft and Coulomb regimes, respectively.
In the top-quark case, we have $\alpha_s=0.146$ at the Coulomb scale and
$\alpha_s=0.196$ at the ultrasoft scale.
Both solutions are in the perturbative region.
In the bottom-quark case, we obtain $\alpha_s=0.34$ at the Coulomb scale and
$\alpha_s=0.47$ at the ultrasoft scale.
The last value seems to be too large for a reliable perturbative calculation.
Thus, to be on the safe side, we redefine the ultrasoft scale to be
$3\alpha_s^2m_b$.
This leads to $\alpha_s=0.34$, which coincides with the value at the Coulomb
scale. 
The $\mu_f$ dependence is cancelled in the total result by hard terms of the
form $\ln(\mu_f/m_q)$ and by soft terms of the form $\ln(\mu_f/\alpha_sm_q)$.
In order to minimize these unknown logarithmic contributions and to absorb the
large logarithms into the know ultrasoft terms, we select $\mu_f$ from the
interval $\alpha_sm_q<\mu_f<m_q$.  

From Figs.~2a and 3a, we observe that, depending on $\mu_f$, the ultrasoft
N$^3$LO corrections to the energy levels of the $b\bar b$ and $t\bar t$
systems may be as large as $-120$~MeV and $+120$~MeV, respectively.
As we see from Figs.~2b and 3b, the ultrasoft N$^3$LO wave-function correction
reaches $-40\%$ for the $b\bar b$ system and $-7\%$ for the $t\bar t$ system.
The circumstance that the corrections are close to zero at some points of the
interval $\alpha_sm_q<\mu_f<m_q$ justifies this choice of factorization scale. 
It is interesting to compare the ultrasoft N$^3$LO corrections presented in
Figs.~2 and 3 with the corresponding NLO and NNLO corrections, which we 
extract from Ref.~\cite{PP}.
In the bottom-quark case, the NLO (NNLO) contributions to $\Delta E_1$ and
$\Delta\psi_1^2$ approximately amount to $-100$~MeV ($-200$~MeV) and
$+50\%$ ($+150\%$), respectively.
In the top-quark case, the corresponding values are $-700$~MeV ($-700$~MeV)
and $-15\%$ ($+15\%$), respectively.
This comparison nicely illustrates the numerical significance of the new
N$^3$LO corrections.

Interesting, relevant and timely applications of our results include the
studies of top-quark pair production at threshold, of low-lying $\Upsilon$
resonances and of $\Upsilon$ sum rules.
In the case of the 1S $\Upsilon$ resonance, where local duality is expected to
work, it is interesting to compare the perturbative ultrasoft contribution,
related to the scale $\alpha_s^2m_b\approx1$~GeV, with the leading
nonperturbative contribution of the gluon condensate, due to nonperturbative
fluctuations at the scale $\Lambda_{\rm QCD}$, which originates from a diagram
of the type shown in Fig.~1 with a broken gluon propagator.
In the case of the 1S energy level, the latter is given by \cite{VolLeu}
\begin{equation}
\Delta E_1=\frac{117m_q}{1275\lambda_s^4}
\left\langle\alpha_s G^a_{\mu\nu} G^{a\mu\nu}\right\rangle.
\end{equation}
Using $\langle\alpha_sG^2\rangle=0.06$~GeV$^4$, we have
$\Delta E_1\approx60$~MeV, which is comparable to the ultrasoft contribution.

In the case of the $J/\psi$ resonance, which is sometimes optimistically
considered to be a Coulomb system, the ultrasoft scale is of the order of
$\Lambda_{\rm QCD}$ if one assumes $\alpha_s$ to freeze at 1~GeV.
Thus, one has to accurately separate perturbative and nonperturbative
contributions.
Nevertheless, our results suggest that the $J/\psi$ resonance cannot be
regarded as a Coulomb potential system.
Indeed, the potential model can be destroyed either by the large contribution
from the gluon condensate, which is nonpotential because the nonperturbative
scale $\Lambda_{\rm QCD}$ is below the potential scale
$\alpha_sm_c\approx700$~MeV, or by the large perturbative retardation
contribution.
The first one is proportional to $1/\alpha_s^4$, while the second one is
proportional to $\alpha_s^5$.
We observe that, due to their different dependence on $\alpha_s$, these two
contributions cannot be simultaneously small for any low-scale evolution of
the coupling constant.
For $m_c\approx1.4$~GeV and $\alpha_s\approx0.5$, the typical size of
nonpotential contributions to the $J/\psi$ energy levels is about 400~MeV,
which is comparable to the inverse Bohr radius.
This clearly indicates that the $J/\psi$ system is far from being Coulombic. 

\section{Conclusions}

We conclude with a few general remarks concerning the structure of
higher-order corrections in pNRQCD.
At NLO, the only source of corrections is the running of $\alpha_s$.
At NNLO, higher-dimensional operators start to contribute.
At N$^3$LO, retardation effects, which cannot be described by local operators,
enter the game.
The leading retardation effects, which are under consideration here, arise
from the chromoelectric dipole interaction of heavy quarkonium with virtual
ultrasoft gluons as depicted in Fig.~1.
To our knowledge, these effects have not been studied elsewhere in the 
literature.
We emphasize that they constitute a genuinely new feature, which is absent in
NLO and NNLO.
In particular, they are not contained in any of the popular renormalon-based
mass definitions.  
On the other hand, they are expected to be the last source of unexpectedly
large corrections.
Thus, our result sets the scale of the N$^3$LO corrections.

In this paper, we took a first step towards the N$^3$LO analysis of the
heavy-quark threshold dynamics.
Our results entirely account for the ultrasoft-scale physics and should be
complemented by the soft and hard contributions. 
The analysis of the last one includes the calculations of the three-loop hard
matching coefficient and the three-loop potential.
The corresponding two-loop results may be found in Refs.~\cite{CzaBen,PetSch},
respectively.
Some N$^3$LO results on the effective Lagrangian were presented in
Ref.~\cite{ManPin}.

\vspace{1cm}
\noindent
{\bf Acknowledgements}
\smallskip

\noindent
B.A.K. thanks the Theory Group of the Werner-Heisenberg-Institut for the
hospitality extended to him during a visit when this paper was finalized.
A.A.P. gratefully acknowledges discussions with K. Melnikov.
The work of A.A.P. is supported in part
by the Volkswagen Foundation under Contract No.~I/73611,
by the Russian Academy of Sciences through Grant No.~37 and
by and Russian Fund for Basic Research under Contract No.~97-02-17065.

\newpage

\section*{Appendix}

Only the S-wave component of the (colour-singlet) Green function, with
angular-momen\-tum quantum number $l=0$, contributes to its value at the
origin and, therefore, to the leading ultrasoft corrections to the
vacuum-polarization function. 
This means that only the $l=1$ component of the colour-octet wave function
contributes to the matrix element $\langle{\bf r}\rangle_{{\bf k}n}$.
The corresponding Coulomb wave functions for the attractive (singlet) and
repulsive (octet) potentials read
\begin{eqnarray}
\psi_{n0}(r)&=&{1\over\sqrt{4\pi}}R_{n0}(r)\,,
\nonumber\\
\psi^o_{{\bf k}1}({\bf r})&=&e^{i(\pi/2-\delta^C_1)}{3\over 2k}R_{k1}(r)
\frac{{\bf k}\cdot{\bf r}}{kr}\,,
\end{eqnarray}
where $\delta^C_1$ is the $l=1$ Coulomb phase and
\begin{eqnarray}
R_{n0}(r)&=&2\left({\lambda_s\over n}\right)^{3/2}e^{-\lambda_sr/n}
F\left(1-n,2,\frac{2\lambda_sr}{n}\right)\,,
\nonumber\\
R_{k1}(r)&=&{\sqrt{8\pi}\over 3}kr
\left({\nu^2+1\over\nu(e^{2\pi\nu}-1)}\right)^{1/2}e^{ikr}F(2+i\nu,4,-i2kr)\,.
\end{eqnarray}
Here, $F$ is the confluent hypergeometric function, $\nu=\lambda_s\rho_1/k$ and
\begin{equation}
\rho_n={1\over n}\left({C_A\over 2C_F}-1\right)={1\over 8n}\,.
\end{equation}
After some algebra, we obtain the QCD Bethe logarithms of Eq.~(\ref{eq:be})
in terms of one-parameter integrals of elementary functions, as
\begin{equation}
L^{E,\psi}_n=\int_0^\infty d\nu Y_n^{E,\psi}(\nu)X^2_n(\nu)\,,
\end{equation}
where
\begin{eqnarray}
Y^E_n(\nu)&=&{2^6\rho_n^5\nu(\nu^2+1)e^{4\nu\arctan\nu/\rho_n}\over 
n^2(\nu^2+\rho_n^2)^3(e^{2\pi\nu}-1)}\ln{n^2\nu^2\over \nu^2+\rho_n^2}\,,
\nonumber\\
Y^\psi_n(\nu)&=&{\nu^2\over(\nu^2+\rho_n^2)}Y^E_n(\nu)\,,
\end{eqnarray}
and
\begin{eqnarray}
X_1(\nu)&=&\rho_1+2\,,
\nonumber\\
X_2(\nu)&=&{\nu^2(2\rho_2^2+9\rho_2+8)-\rho_2^2(\rho_2+4)\over
(\nu^2+\rho_2^2)}\,,
\nonumber\\
X_3(\nu)&=&{\nu^4(8\rho_3^3+60\rho_3^2+123\rho_3+66)
-2\nu^2\rho_3^2(6\rho_3^2+41\rho_3+54)+3\rho_3^4(\rho_3+6)\over
3(\nu^2+\rho_3^2)^2}\,.
\end{eqnarray}
Since it is usually sufficient to consider $n=1,2,3$ in practical applications
and the expressions for $X_n$ with $n>3$ are somewhat cumbersome, we refrain 
to listing the latter.

\newpage

\section*{Figure captions}

\noindent
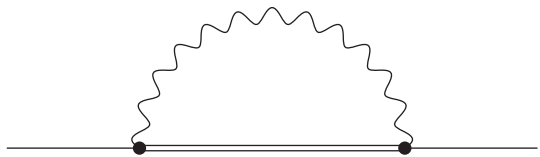
{\bf Fig. 1.} Feynman diagram giving rise to the ultrasoft contribution at
N$^3$LO.
The single and double lines stand for the singlet and octet Green functions,
respectively, the wavy line represents the ultrasoft-gluon propagator in the 
Coulomb gauge, and the vertices correspond to the chromoelectric dipole
interaction. 

\smallskip

\noindent
{\bf Fig. 2.} Ultrasoft corrections to (a) the energy level $E_1$ and (b) the
square of the wave function at the origin $|\psi_1(0)|^2$ as functions of
the factorization scale $\mu_f$ for the $b\bar b$ ground state, with principal
quantum number $n=1$.

\smallskip

\noindent
{\bf Fig. 3.} Same as in Fig.~2, but for $t\bar t$.

\bigskip

\begin{center}

\begin{picture}(300,56)(0,0)
\Line(50,10)(100,10)
\Line(200,10)(250,10)
\Line(100,11)(200,11)
\Line(100,9)(200,9)
\PhotonArc(150,10)(50,0,180){3}{12.5}      
\Vertex(100,10){2.5} \Vertex(200,10){2.5}
\Text(150,60)[cb]{}
\end{picture}

\smallskip

{\large\bf Fig.~1}

\end{center}

\newpage

\begin{center}

% GNUPLOT: LaTeX picture
\setlength{\unitlength}{0.240900pt}
\ifx\plotpoint\undefined\newsavebox{\plotpoint}\fi
\sbox{\plotpoint}{\rule[-0.200pt]{0.400pt}{0.400pt}}%
\begin{picture}(1500,900)(0,0)
\font\gnuplot=cmr10 at 10pt
\gnuplot
\sbox{\plotpoint}{\rule[-0.200pt]{0.400pt}{0.400pt}}%
\put(221.0,163.0){\rule[-0.200pt]{4.818pt}{0.400pt}}
\put(201,163){\makebox(0,0)[r]{-120}}
\put(1460.0,163.0){\rule[-0.200pt]{4.818pt}{0.400pt}}
\put(221.0,250.0){\rule[-0.200pt]{4.818pt}{0.400pt}}
\put(201,250){\makebox(0,0)[r]{-100}}
\put(1460.0,250.0){\rule[-0.200pt]{4.818pt}{0.400pt}}
\put(221.0,337.0){\rule[-0.200pt]{4.818pt}{0.400pt}}
\put(201,337){\makebox(0,0)[r]{-80}}
\put(1460.0,337.0){\rule[-0.200pt]{4.818pt}{0.400pt}}
\put(221.0,424.0){\rule[-0.200pt]{4.818pt}{0.400pt}}
\put(201,424){\makebox(0,0)[r]{-60}}
\put(1460.0,424.0){\rule[-0.200pt]{4.818pt}{0.400pt}}
\put(221.0,511.0){\rule[-0.200pt]{4.818pt}{0.400pt}}
\put(201,511){\makebox(0,0)[r]{-40}}
\put(1460.0,511.0){\rule[-0.200pt]{4.818pt}{0.400pt}}
\put(221.0,598.0){\rule[-0.200pt]{4.818pt}{0.400pt}}
\put(201,598){\makebox(0,0)[r]{-20}}
\put(1460.0,598.0){\rule[-0.200pt]{4.818pt}{0.400pt}}
\put(221.0,685.0){\rule[-0.200pt]{4.818pt}{0.400pt}}
\put(201,685){\makebox(0,0)[r]{0}}
\put(1460.0,685.0){\rule[-0.200pt]{4.818pt}{0.400pt}}
\put(221.0,772.0){\rule[-0.200pt]{4.818pt}{0.400pt}}
\put(201,772){\makebox(0,0)[r]{20}}
\put(1460.0,772.0){\rule[-0.200pt]{4.818pt}{0.400pt}}
\put(221.0,859.0){\rule[-0.200pt]{4.818pt}{0.400pt}}
\put(201,859){\makebox(0,0)[r]{40}}
\put(1460.0,859.0){\rule[-0.200pt]{4.818pt}{0.400pt}}
\put(221.0,163.0){\rule[-0.200pt]{0.400pt}{4.818pt}}
\put(221,122){\makebox(0,0){1.5}}
\put(221.0,839.0){\rule[-0.200pt]{0.400pt}{4.818pt}}
\put(401.0,163.0){\rule[-0.200pt]{0.400pt}{4.818pt}}
\put(401,122){\makebox(0,0){2}}
\put(401.0,839.0){\rule[-0.200pt]{0.400pt}{4.818pt}}
\put(581.0,163.0){\rule[-0.200pt]{0.400pt}{4.818pt}}
\put(581,122){\makebox(0,0){2.5}}
\put(581.0,839.0){\rule[-0.200pt]{0.400pt}{4.818pt}}
\put(761.0,163.0){\rule[-0.200pt]{0.400pt}{4.818pt}}
\put(761,122){\makebox(0,0){3}}
\put(761.0,839.0){\rule[-0.200pt]{0.400pt}{4.818pt}}
\put(940.0,163.0){\rule[-0.200pt]{0.400pt}{4.818pt}}
\put(940,122){\makebox(0,0){3.5}}
\put(940.0,839.0){\rule[-0.200pt]{0.400pt}{4.818pt}}
\put(1120.0,163.0){\rule[-0.200pt]{0.400pt}{4.818pt}}
\put(1120,122){\makebox(0,0){4}}
\put(1120.0,839.0){\rule[-0.200pt]{0.400pt}{4.818pt}}
\put(1300.0,163.0){\rule[-0.200pt]{0.400pt}{4.818pt}}
\put(1300,122){\makebox(0,0){4.5}}
\put(1300.0,839.0){\rule[-0.200pt]{0.400pt}{4.818pt}}
\put(1480.0,163.0){\rule[-0.200pt]{0.400pt}{4.818pt}}
\put(1480,122){\makebox(0,0){5}}
\put(1480.0,839.0){\rule[-0.200pt]{0.400pt}{4.818pt}}
\put(221.0,685.0){\rule[-0.200pt]{303.293pt}{0.400pt}}
\put(221.0,163.0){\rule[-0.200pt]{303.293pt}{0.400pt}}
\put(1480.0,163.0){\rule[-0.200pt]{0.400pt}{167.666pt}}
\put(221.0,859.0){\rule[-0.200pt]{303.293pt}{0.400pt}}
\put(-10,511){\makebox(0,0){$\Delta E_1$ (MeV)}}
\put(850,21){\makebox(0,0){$\mu_f$ (GeV)}}
\put(221.0,163.0){\rule[-0.200pt]{0.400pt}{167.666pt}}
\put(268,171){\usebox{\plotpoint}}
\multiput(268.00,171.58)(0.513,0.498){67}{\rule{0.511pt}{0.120pt}}
\multiput(268.00,170.17)(34.939,35.000){2}{\rule{0.256pt}{0.400pt}}
\multiput(304.00,206.58)(0.529,0.498){65}{\rule{0.524pt}{0.120pt}}
\multiput(304.00,205.17)(34.913,34.000){2}{\rule{0.262pt}{0.400pt}}
\multiput(340.00,240.58)(0.562,0.497){61}{\rule{0.550pt}{0.120pt}}
\multiput(340.00,239.17)(34.858,32.000){2}{\rule{0.275pt}{0.400pt}}
\multiput(376.00,272.58)(0.600,0.497){57}{\rule{0.580pt}{0.120pt}}
\multiput(376.00,271.17)(34.796,30.000){2}{\rule{0.290pt}{0.400pt}}
\multiput(412.00,302.58)(0.621,0.497){55}{\rule{0.597pt}{0.120pt}}
\multiput(412.00,301.17)(34.762,29.000){2}{\rule{0.298pt}{0.400pt}}
\multiput(448.00,331.58)(0.667,0.497){51}{\rule{0.633pt}{0.120pt}}
\multiput(448.00,330.17)(34.685,27.000){2}{\rule{0.317pt}{0.400pt}}
\multiput(484.00,358.58)(0.693,0.497){49}{\rule{0.654pt}{0.120pt}}
\multiput(484.00,357.17)(34.643,26.000){2}{\rule{0.327pt}{0.400pt}}
\multiput(520.00,384.58)(0.693,0.497){49}{\rule{0.654pt}{0.120pt}}
\multiput(520.00,383.17)(34.643,26.000){2}{\rule{0.327pt}{0.400pt}}
\multiput(556.00,410.58)(0.752,0.496){45}{\rule{0.700pt}{0.120pt}}
\multiput(556.00,409.17)(34.547,24.000){2}{\rule{0.350pt}{0.400pt}}
\multiput(592.00,434.58)(0.785,0.496){43}{\rule{0.726pt}{0.120pt}}
\multiput(592.00,433.17)(34.493,23.000){2}{\rule{0.363pt}{0.400pt}}
\multiput(628.00,457.58)(0.822,0.496){41}{\rule{0.755pt}{0.120pt}}
\multiput(628.00,456.17)(34.434,22.000){2}{\rule{0.377pt}{0.400pt}}
\multiput(664.00,479.58)(0.822,0.496){41}{\rule{0.755pt}{0.120pt}}
\multiput(664.00,478.17)(34.434,22.000){2}{\rule{0.377pt}{0.400pt}}
\multiput(700.00,501.58)(0.905,0.496){37}{\rule{0.820pt}{0.119pt}}
\multiput(700.00,500.17)(34.298,20.000){2}{\rule{0.410pt}{0.400pt}}
\multiput(736.00,521.58)(0.905,0.496){37}{\rule{0.820pt}{0.119pt}}
\multiput(736.00,520.17)(34.298,20.000){2}{\rule{0.410pt}{0.400pt}}
\multiput(772.00,541.58)(0.905,0.496){37}{\rule{0.820pt}{0.119pt}}
\multiput(772.00,540.17)(34.298,20.000){2}{\rule{0.410pt}{0.400pt}}
\multiput(808.00,561.58)(0.954,0.495){35}{\rule{0.858pt}{0.119pt}}
\multiput(808.00,560.17)(34.219,19.000){2}{\rule{0.429pt}{0.400pt}}
\multiput(844.00,580.58)(1.008,0.495){33}{\rule{0.900pt}{0.119pt}}
\multiput(844.00,579.17)(34.132,18.000){2}{\rule{0.450pt}{0.400pt}}
\multiput(880.00,598.58)(1.008,0.495){33}{\rule{0.900pt}{0.119pt}}
\multiput(880.00,597.17)(34.132,18.000){2}{\rule{0.450pt}{0.400pt}}
\multiput(916.00,616.58)(1.069,0.495){31}{\rule{0.947pt}{0.119pt}}
\multiput(916.00,615.17)(34.034,17.000){2}{\rule{0.474pt}{0.400pt}}
\multiput(952.00,633.58)(1.069,0.495){31}{\rule{0.947pt}{0.119pt}}
\multiput(952.00,632.17)(34.034,17.000){2}{\rule{0.474pt}{0.400pt}}
\multiput(988.00,650.58)(1.137,0.494){29}{\rule{1.000pt}{0.119pt}}
\multiput(988.00,649.17)(33.924,16.000){2}{\rule{0.500pt}{0.400pt}}
\multiput(1024.00,666.58)(1.137,0.494){29}{\rule{1.000pt}{0.119pt}}
\multiput(1024.00,665.17)(33.924,16.000){2}{\rule{0.500pt}{0.400pt}}
\multiput(1060.00,682.58)(1.215,0.494){27}{\rule{1.060pt}{0.119pt}}
\multiput(1060.00,681.17)(33.800,15.000){2}{\rule{0.530pt}{0.400pt}}
\multiput(1096.00,697.58)(1.215,0.494){27}{\rule{1.060pt}{0.119pt}}
\multiput(1096.00,696.17)(33.800,15.000){2}{\rule{0.530pt}{0.400pt}}
\multiput(1132.00,712.58)(1.215,0.494){27}{\rule{1.060pt}{0.119pt}}
\multiput(1132.00,711.17)(33.800,15.000){2}{\rule{0.530pt}{0.400pt}}
\multiput(1168.00,727.58)(1.305,0.494){25}{\rule{1.129pt}{0.119pt}}
\multiput(1168.00,726.17)(33.658,14.000){2}{\rule{0.564pt}{0.400pt}}
\multiput(1204.00,741.58)(1.305,0.494){25}{\rule{1.129pt}{0.119pt}}
\multiput(1204.00,740.17)(33.658,14.000){2}{\rule{0.564pt}{0.400pt}}
\multiput(1240.00,755.58)(1.305,0.494){25}{\rule{1.129pt}{0.119pt}}
\multiput(1240.00,754.17)(33.658,14.000){2}{\rule{0.564pt}{0.400pt}}
\multiput(1276.00,769.58)(1.408,0.493){23}{\rule{1.208pt}{0.119pt}}
\multiput(1276.00,768.17)(33.493,13.000){2}{\rule{0.604pt}{0.400pt}}
\multiput(1312.00,782.58)(1.408,0.493){23}{\rule{1.208pt}{0.119pt}}
\multiput(1312.00,781.17)(33.493,13.000){2}{\rule{0.604pt}{0.400pt}}
\multiput(1348.00,795.58)(1.408,0.493){23}{\rule{1.208pt}{0.119pt}}
\multiput(1348.00,794.17)(33.493,13.000){2}{\rule{0.604pt}{0.400pt}}
\end{picture}

\smallskip

{\large\bf Fig.~2a}

\bigskip

% GNUPLOT: LaTeX picture
\setlength{\unitlength}{0.240900pt}
\ifx\plotpoint\undefined\newsavebox{\plotpoint}\fi
\begin{picture}(1500,900)(0,0)
\font\gnuplot=cmr10 at 10pt
\gnuplot
\sbox{\plotpoint}{\rule[-0.200pt]{0.400pt}{0.400pt}}%
\put(221.0,163.0){\rule[-0.200pt]{4.818pt}{0.400pt}}
\put(201,163){\makebox(0,0)[r]{-40}}
\put(1460.0,163.0){\rule[-0.200pt]{4.818pt}{0.400pt}}
\put(221.0,279.0){\rule[-0.200pt]{4.818pt}{0.400pt}}
\put(201,279){\makebox(0,0)[r]{-35}}
\put(1460.0,279.0){\rule[-0.200pt]{4.818pt}{0.400pt}}
\put(221.0,395.0){\rule[-0.200pt]{4.818pt}{0.400pt}}
\put(201,395){\makebox(0,0)[r]{-30}}
\put(1460.0,395.0){\rule[-0.200pt]{4.818pt}{0.400pt}}
\put(221.0,511.0){\rule[-0.200pt]{4.818pt}{0.400pt}}
\put(201,511){\makebox(0,0)[r]{-25}}
\put(1460.0,511.0){\rule[-0.200pt]{4.818pt}{0.400pt}}
\put(221.0,627.0){\rule[-0.200pt]{4.818pt}{0.400pt}}
\put(201,627){\makebox(0,0)[r]{-20}}
\put(1460.0,627.0){\rule[-0.200pt]{4.818pt}{0.400pt}}
\put(221.0,743.0){\rule[-0.200pt]{4.818pt}{0.400pt}}
\put(201,743){\makebox(0,0)[r]{-15}}
\put(1460.0,743.0){\rule[-0.200pt]{4.818pt}{0.400pt}}
\put(221.0,859.0){\rule[-0.200pt]{4.818pt}{0.400pt}}
\put(201,859){\makebox(0,0)[r]{-10}}
\put(1460.0,859.0){\rule[-0.200pt]{4.818pt}{0.400pt}}
\put(221.0,163.0){\rule[-0.200pt]{0.400pt}{4.818pt}}
\put(221,122){\makebox(0,0){1.5}}
\put(221.0,839.0){\rule[-0.200pt]{0.400pt}{4.818pt}}
\put(401.0,163.0){\rule[-0.200pt]{0.400pt}{4.818pt}}
\put(401,122){\makebox(0,0){2}}
\put(401.0,839.0){\rule[-0.200pt]{0.400pt}{4.818pt}}
\put(581.0,163.0){\rule[-0.200pt]{0.400pt}{4.818pt}}
\put(581,122){\makebox(0,0){2.5}}
\put(581.0,839.0){\rule[-0.200pt]{0.400pt}{4.818pt}}
\put(761.0,163.0){\rule[-0.200pt]{0.400pt}{4.818pt}}
\put(761,122){\makebox(0,0){3}}
\put(761.0,839.0){\rule[-0.200pt]{0.400pt}{4.818pt}}
\put(940.0,163.0){\rule[-0.200pt]{0.400pt}{4.818pt}}
\put(940,122){\makebox(0,0){3.5}}
\put(940.0,839.0){\rule[-0.200pt]{0.400pt}{4.818pt}}
\put(1120.0,163.0){\rule[-0.200pt]{0.400pt}{4.818pt}}
\put(1120,122){\makebox(0,0){4}}
\put(1120.0,839.0){\rule[-0.200pt]{0.400pt}{4.818pt}}
\put(1300.0,163.0){\rule[-0.200pt]{0.400pt}{4.818pt}}
\put(1300,122){\makebox(0,0){4.5}}
\put(1300.0,839.0){\rule[-0.200pt]{0.400pt}{4.818pt}}
\put(1480.0,163.0){\rule[-0.200pt]{0.400pt}{4.818pt}}
\put(1480,122){\makebox(0,0){5}}
\put(1480.0,839.0){\rule[-0.200pt]{0.400pt}{4.818pt}}
\put(221.0,163.0){\rule[-0.200pt]{303.293pt}{0.400pt}}
\put(1480.0,163.0){\rule[-0.200pt]{0.400pt}{167.666pt}}
\put(221.0,859.0){\rule[-0.200pt]{303.293pt}{0.400pt}}
\put(1,511){\makebox(0,0){$\Delta \psi^2_1\times 10^2$}}
\put(850,21){\makebox(0,0){$\mu_f$ (GeV)}}
\put(221.0,163.0){\rule[-0.200pt]{0.400pt}{167.666pt}}
\put(268,830){\usebox{\plotpoint}}
\multiput(268.58,827.88)(0.498,-0.513){69}{\rule{0.120pt}{0.511pt}}
\multiput(267.17,828.94)(36.000,-35.939){2}{\rule{0.400pt}{0.256pt}}
\multiput(304.00,791.92)(0.529,-0.498){65}{\rule{0.524pt}{0.120pt}}
\multiput(304.00,792.17)(34.913,-34.000){2}{\rule{0.262pt}{0.400pt}}
\multiput(340.00,757.92)(0.545,-0.497){63}{\rule{0.536pt}{0.120pt}}
\multiput(340.00,758.17)(34.887,-33.000){2}{\rule{0.268pt}{0.400pt}}
\multiput(376.00,724.92)(0.562,-0.497){61}{\rule{0.550pt}{0.120pt}}
\multiput(376.00,725.17)(34.858,-32.000){2}{\rule{0.275pt}{0.400pt}}
\multiput(412.00,692.92)(0.600,-0.497){57}{\rule{0.580pt}{0.120pt}}
\multiput(412.00,693.17)(34.796,-30.000){2}{\rule{0.290pt}{0.400pt}}
\multiput(448.00,662.92)(0.643,-0.497){53}{\rule{0.614pt}{0.120pt}}
\multiput(448.00,663.17)(34.725,-28.000){2}{\rule{0.307pt}{0.400pt}}
\multiput(484.00,634.92)(0.667,-0.497){51}{\rule{0.633pt}{0.120pt}}
\multiput(484.00,635.17)(34.685,-27.000){2}{\rule{0.317pt}{0.400pt}}
\multiput(520.00,607.92)(0.693,-0.497){49}{\rule{0.654pt}{0.120pt}}
\multiput(520.00,608.17)(34.643,-26.000){2}{\rule{0.327pt}{0.400pt}}
\multiput(556.00,581.92)(0.722,-0.497){47}{\rule{0.676pt}{0.120pt}}
\multiput(556.00,582.17)(34.597,-25.000){2}{\rule{0.338pt}{0.400pt}}
\multiput(592.00,556.92)(0.722,-0.497){47}{\rule{0.676pt}{0.120pt}}
\multiput(592.00,557.17)(34.597,-25.000){2}{\rule{0.338pt}{0.400pt}}
\multiput(628.00,531.92)(0.785,-0.496){43}{\rule{0.726pt}{0.120pt}}
\multiput(628.00,532.17)(34.493,-23.000){2}{\rule{0.363pt}{0.400pt}}
\multiput(664.00,508.92)(0.822,-0.496){41}{\rule{0.755pt}{0.120pt}}
\multiput(664.00,509.17)(34.434,-22.000){2}{\rule{0.377pt}{0.400pt}}
\multiput(700.00,486.92)(0.822,-0.496){41}{\rule{0.755pt}{0.120pt}}
\multiput(700.00,487.17)(34.434,-22.000){2}{\rule{0.377pt}{0.400pt}}
\multiput(736.00,464.92)(0.905,-0.496){37}{\rule{0.820pt}{0.119pt}}
\multiput(736.00,465.17)(34.298,-20.000){2}{\rule{0.410pt}{0.400pt}}
\multiput(772.00,444.92)(0.862,-0.496){39}{\rule{0.786pt}{0.119pt}}
\multiput(772.00,445.17)(34.369,-21.000){2}{\rule{0.393pt}{0.400pt}}
\multiput(808.00,423.92)(0.954,-0.495){35}{\rule{0.858pt}{0.119pt}}
\multiput(808.00,424.17)(34.219,-19.000){2}{\rule{0.429pt}{0.400pt}}
\multiput(844.00,404.92)(0.954,-0.495){35}{\rule{0.858pt}{0.119pt}}
\multiput(844.00,405.17)(34.219,-19.000){2}{\rule{0.429pt}{0.400pt}}
\multiput(880.00,385.92)(1.008,-0.495){33}{\rule{0.900pt}{0.119pt}}
\multiput(880.00,386.17)(34.132,-18.000){2}{\rule{0.450pt}{0.400pt}}
\multiput(916.00,367.92)(1.008,-0.495){33}{\rule{0.900pt}{0.119pt}}
\multiput(916.00,368.17)(34.132,-18.000){2}{\rule{0.450pt}{0.400pt}}
\multiput(952.00,349.92)(1.008,-0.495){33}{\rule{0.900pt}{0.119pt}}
\multiput(952.00,350.17)(34.132,-18.000){2}{\rule{0.450pt}{0.400pt}}
\multiput(988.00,331.92)(1.069,-0.495){31}{\rule{0.947pt}{0.119pt}}
\multiput(988.00,332.17)(34.034,-17.000){2}{\rule{0.474pt}{0.400pt}}
\multiput(1024.00,314.92)(1.137,-0.494){29}{\rule{1.000pt}{0.119pt}}
\multiput(1024.00,315.17)(33.924,-16.000){2}{\rule{0.500pt}{0.400pt}}
\multiput(1060.00,298.92)(1.137,-0.494){29}{\rule{1.000pt}{0.119pt}}
\multiput(1060.00,299.17)(33.924,-16.000){2}{\rule{0.500pt}{0.400pt}}
\multiput(1096.00,282.92)(1.137,-0.494){29}{\rule{1.000pt}{0.119pt}}
\multiput(1096.00,283.17)(33.924,-16.000){2}{\rule{0.500pt}{0.400pt}}
\multiput(1132.00,266.92)(1.215,-0.494){27}{\rule{1.060pt}{0.119pt}}
\multiput(1132.00,267.17)(33.800,-15.000){2}{\rule{0.530pt}{0.400pt}}
\multiput(1168.00,251.92)(1.215,-0.494){27}{\rule{1.060pt}{0.119pt}}
\multiput(1168.00,252.17)(33.800,-15.000){2}{\rule{0.530pt}{0.400pt}}
\multiput(1204.00,236.92)(1.305,-0.494){25}{\rule{1.129pt}{0.119pt}}
\multiput(1204.00,237.17)(33.658,-14.000){2}{\rule{0.564pt}{0.400pt}}
\multiput(1240.00,222.92)(1.305,-0.494){25}{\rule{1.129pt}{0.119pt}}
\multiput(1240.00,223.17)(33.658,-14.000){2}{\rule{0.564pt}{0.400pt}}
\multiput(1276.00,208.92)(1.305,-0.494){25}{\rule{1.129pt}{0.119pt}}
\multiput(1276.00,209.17)(33.658,-14.000){2}{\rule{0.564pt}{0.400pt}}
\multiput(1312.00,194.92)(1.305,-0.494){25}{\rule{1.129pt}{0.119pt}}
\multiput(1312.00,195.17)(33.658,-14.000){2}{\rule{0.564pt}{0.400pt}}
\multiput(1348.00,180.92)(1.408,-0.493){23}{\rule{1.208pt}{0.119pt}}
\multiput(1348.00,181.17)(33.493,-13.000){2}{\rule{0.604pt}{0.400pt}}
\end{picture}

\smallskip

{\large \bf Fig.~2b}

\newpage

% GNUPLOT: LaTeX picture
\setlength{\unitlength}{0.240900pt}
\ifx\plotpoint\undefined\newsavebox{\plotpoint}\fi
\begin{picture}(1500,900)(0,0)
\font\gnuplot=cmr10 at 10pt
\gnuplot
\sbox{\plotpoint}{\rule[-0.200pt]{0.400pt}{0.400pt}}%
\put(221.0,163.0){\rule[-0.200pt]{4.818pt}{0.400pt}}
\put(201,163){\makebox(0,0)[r]{-60}}
\put(1460.0,163.0){\rule[-0.200pt]{4.818pt}{0.400pt}}
\put(221.0,233.0){\rule[-0.200pt]{4.818pt}{0.400pt}}
\put(201,233){\makebox(0,0)[r]{-40}}
\put(1460.0,233.0){\rule[-0.200pt]{4.818pt}{0.400pt}}
\put(221.0,302.0){\rule[-0.200pt]{4.818pt}{0.400pt}}
\put(201,302){\makebox(0,0)[r]{-20}}
\put(1460.0,302.0){\rule[-0.200pt]{4.818pt}{0.400pt}}
\put(221.0,372.0){\rule[-0.200pt]{4.818pt}{0.400pt}}
\put(201,372){\makebox(0,0)[r]{0}}
\put(1460.0,372.0){\rule[-0.200pt]{4.818pt}{0.400pt}}
\put(221.0,441.0){\rule[-0.200pt]{4.818pt}{0.400pt}}
\put(201,441){\makebox(0,0)[r]{20}}
\put(1460.0,441.0){\rule[-0.200pt]{4.818pt}{0.400pt}}
\put(221.0,511.0){\rule[-0.200pt]{4.818pt}{0.400pt}}
\put(201,511){\makebox(0,0)[r]{40}}
\put(1460.0,511.0){\rule[-0.200pt]{4.818pt}{0.400pt}}
\put(221.0,581.0){\rule[-0.200pt]{4.818pt}{0.400pt}}
\put(201,581){\makebox(0,0)[r]{60}}
\put(1460.0,581.0){\rule[-0.200pt]{4.818pt}{0.400pt}}
\put(221.0,650.0){\rule[-0.200pt]{4.818pt}{0.400pt}}
\put(201,650){\makebox(0,0)[r]{80}}
\put(1460.0,650.0){\rule[-0.200pt]{4.818pt}{0.400pt}}
\put(221.0,720.0){\rule[-0.200pt]{4.818pt}{0.400pt}}
\put(201,720){\makebox(0,0)[r]{100}}
\put(1460.0,720.0){\rule[-0.200pt]{4.818pt}{0.400pt}}
\put(221.0,789.0){\rule[-0.200pt]{4.818pt}{0.400pt}}
\put(201,789){\makebox(0,0)[r]{120}}
\put(1460.0,789.0){\rule[-0.200pt]{4.818pt}{0.400pt}}
\put(221.0,859.0){\rule[-0.200pt]{4.818pt}{0.400pt}}
\put(201,859){\makebox(0,0)[r]{140}}
\put(1460.0,859.0){\rule[-0.200pt]{4.818pt}{0.400pt}}
\put(221.0,163.0){\rule[-0.200pt]{0.400pt}{4.818pt}}
\put(221,122){\makebox(0,0){20}}
\put(221.0,839.0){\rule[-0.200pt]{0.400pt}{4.818pt}}
\put(378.0,163.0){\rule[-0.200pt]{0.400pt}{4.818pt}}
\put(378,122){\makebox(0,0){40}}
\put(378.0,839.0){\rule[-0.200pt]{0.400pt}{4.818pt}}
\put(536.0,163.0){\rule[-0.200pt]{0.400pt}{4.818pt}}
\put(536,122){\makebox(0,0){60}}
\put(536.0,839.0){\rule[-0.200pt]{0.400pt}{4.818pt}}
\put(693.0,163.0){\rule[-0.200pt]{0.400pt}{4.818pt}}
\put(693,122){\makebox(0,0){80}}
\put(693.0,839.0){\rule[-0.200pt]{0.400pt}{4.818pt}}
\put(851.0,163.0){\rule[-0.200pt]{0.400pt}{4.818pt}}
\put(851,122){\makebox(0,0){100}}
\put(851.0,839.0){\rule[-0.200pt]{0.400pt}{4.818pt}}
\put(1008.0,163.0){\rule[-0.200pt]{0.400pt}{4.818pt}}
\put(1008,122){\makebox(0,0){120}}
\put(1008.0,839.0){\rule[-0.200pt]{0.400pt}{4.818pt}}
\put(1165.0,163.0){\rule[-0.200pt]{0.400pt}{4.818pt}}
\put(1165,122){\makebox(0,0){140}}
\put(1165.0,839.0){\rule[-0.200pt]{0.400pt}{4.818pt}}
\put(1323.0,163.0){\rule[-0.200pt]{0.400pt}{4.818pt}}
\put(1323,122){\makebox(0,0){160}}
\put(1323.0,839.0){\rule[-0.200pt]{0.400pt}{4.818pt}}
\put(1480.0,163.0){\rule[-0.200pt]{0.400pt}{4.818pt}}
\put(1480,122){\makebox(0,0){180}}
\put(1480.0,839.0){\rule[-0.200pt]{0.400pt}{4.818pt}}
\put(221.0,372.0){\rule[-0.200pt]{303.293pt}{0.400pt}}
\put(221.0,163.0){\rule[-0.200pt]{303.293pt}{0.400pt}}
\put(1480.0,163.0){\rule[-0.200pt]{0.400pt}{167.666pt}}
\put(221.0,859.0){\rule[-0.200pt]{303.293pt}{0.400pt}}
\put(-10,511){\makebox(0,0){$\Delta E_1$ (MeV)}}
\put(850,21){\makebox(0,0){$\mu_f$ (GeV)}}
\put(221.0,163.0){\rule[-0.200pt]{0.400pt}{167.666pt}}
\put(265,166){\usebox{\plotpoint}}
\multiput(265.58,166.00)(0.497,0.792){59}{\rule{0.120pt}{0.732pt}}
\multiput(264.17,166.00)(31.000,47.480){2}{\rule{0.400pt}{0.366pt}}
\multiput(296.58,215.00)(0.497,0.688){61}{\rule{0.120pt}{0.650pt}}
\multiput(295.17,215.00)(32.000,42.651){2}{\rule{0.400pt}{0.325pt}}
\multiput(328.58,259.00)(0.497,0.613){59}{\rule{0.120pt}{0.590pt}}
\multiput(327.17,259.00)(31.000,36.775){2}{\rule{0.400pt}{0.295pt}}
\multiput(359.58,297.00)(0.497,0.546){61}{\rule{0.120pt}{0.538pt}}
\multiput(358.17,297.00)(32.000,33.884){2}{\rule{0.400pt}{0.269pt}}
\multiput(391.00,332.58)(0.499,0.497){59}{\rule{0.500pt}{0.120pt}}
\multiput(391.00,331.17)(29.962,31.000){2}{\rule{0.250pt}{0.400pt}}
\multiput(422.00,363.58)(0.551,0.497){55}{\rule{0.541pt}{0.120pt}}
\multiput(422.00,362.17)(30.876,29.000){2}{\rule{0.271pt}{0.400pt}}
\multiput(454.00,392.58)(0.574,0.497){51}{\rule{0.559pt}{0.120pt}}
\multiput(454.00,391.17)(29.839,27.000){2}{\rule{0.280pt}{0.400pt}}
\multiput(485.00,419.58)(0.646,0.496){45}{\rule{0.617pt}{0.120pt}}
\multiput(485.00,418.17)(29.720,24.000){2}{\rule{0.308pt}{0.400pt}}
\multiput(516.00,443.58)(0.697,0.496){43}{\rule{0.657pt}{0.120pt}}
\multiput(516.00,442.17)(30.637,23.000){2}{\rule{0.328pt}{0.400pt}}
\multiput(548.00,466.58)(0.706,0.496){41}{\rule{0.664pt}{0.120pt}}
\multiput(548.00,465.17)(29.623,22.000){2}{\rule{0.332pt}{0.400pt}}
\multiput(579.00,488.58)(0.804,0.496){37}{\rule{0.740pt}{0.119pt}}
\multiput(579.00,487.17)(30.464,20.000){2}{\rule{0.370pt}{0.400pt}}
\multiput(611.00,508.58)(0.820,0.495){35}{\rule{0.753pt}{0.119pt}}
\multiput(611.00,507.17)(29.438,19.000){2}{\rule{0.376pt}{0.400pt}}
\multiput(642.00,527.58)(0.895,0.495){33}{\rule{0.811pt}{0.119pt}}
\multiput(642.00,526.17)(30.316,18.000){2}{\rule{0.406pt}{0.400pt}}
\multiput(674.00,545.58)(0.866,0.495){33}{\rule{0.789pt}{0.119pt}}
\multiput(674.00,544.17)(29.363,18.000){2}{\rule{0.394pt}{0.400pt}}
\multiput(705.00,563.58)(1.009,0.494){29}{\rule{0.900pt}{0.119pt}}
\multiput(705.00,562.17)(30.132,16.000){2}{\rule{0.450pt}{0.400pt}}
\multiput(737.00,579.58)(0.977,0.494){29}{\rule{0.875pt}{0.119pt}}
\multiput(737.00,578.17)(29.184,16.000){2}{\rule{0.438pt}{0.400pt}}
\multiput(768.00,595.58)(1.079,0.494){27}{\rule{0.953pt}{0.119pt}}
\multiput(768.00,594.17)(30.021,15.000){2}{\rule{0.477pt}{0.400pt}}
\multiput(800.00,610.58)(1.121,0.494){25}{\rule{0.986pt}{0.119pt}}
\multiput(800.00,609.17)(28.954,14.000){2}{\rule{0.493pt}{0.400pt}}
\multiput(831.00,624.58)(1.158,0.494){25}{\rule{1.014pt}{0.119pt}}
\multiput(831.00,623.17)(29.895,14.000){2}{\rule{0.507pt}{0.400pt}}
\multiput(863.00,638.58)(1.210,0.493){23}{\rule{1.054pt}{0.119pt}}
\multiput(863.00,637.17)(28.813,13.000){2}{\rule{0.527pt}{0.400pt}}
\multiput(894.00,651.58)(1.250,0.493){23}{\rule{1.085pt}{0.119pt}}
\multiput(894.00,650.17)(29.749,13.000){2}{\rule{0.542pt}{0.400pt}}
\multiput(926.00,664.58)(1.315,0.492){21}{\rule{1.133pt}{0.119pt}}
\multiput(926.00,663.17)(28.648,12.000){2}{\rule{0.567pt}{0.400pt}}
\multiput(957.00,676.58)(1.358,0.492){21}{\rule{1.167pt}{0.119pt}}
\multiput(957.00,675.17)(29.579,12.000){2}{\rule{0.583pt}{0.400pt}}
\multiput(989.00,688.58)(1.439,0.492){19}{\rule{1.227pt}{0.118pt}}
\multiput(989.00,687.17)(28.453,11.000){2}{\rule{0.614pt}{0.400pt}}
\multiput(1020.00,699.58)(1.486,0.492){19}{\rule{1.264pt}{0.118pt}}
\multiput(1020.00,698.17)(29.377,11.000){2}{\rule{0.632pt}{0.400pt}}
\multiput(1052.00,710.58)(1.439,0.492){19}{\rule{1.227pt}{0.118pt}}
\multiput(1052.00,709.17)(28.453,11.000){2}{\rule{0.614pt}{0.400pt}}
\multiput(1083.00,721.58)(1.590,0.491){17}{\rule{1.340pt}{0.118pt}}
\multiput(1083.00,720.17)(28.219,10.000){2}{\rule{0.670pt}{0.400pt}}
\multiput(1114.00,731.58)(1.486,0.492){19}{\rule{1.264pt}{0.118pt}}
\multiput(1114.00,730.17)(29.377,11.000){2}{\rule{0.632pt}{0.400pt}}
\multiput(1146.00,742.59)(1.776,0.489){15}{\rule{1.478pt}{0.118pt}}
\multiput(1146.00,741.17)(27.933,9.000){2}{\rule{0.739pt}{0.400pt}}
\multiput(1177.00,751.58)(1.642,0.491){17}{\rule{1.380pt}{0.118pt}}
\multiput(1177.00,750.17)(29.136,10.000){2}{\rule{0.690pt}{0.400pt}}
\multiput(1209.00,761.59)(1.776,0.489){15}{\rule{1.478pt}{0.118pt}}
\multiput(1209.00,760.17)(27.933,9.000){2}{\rule{0.739pt}{0.400pt}}
\multiput(1240.00,770.59)(1.834,0.489){15}{\rule{1.522pt}{0.118pt}}
\multiput(1240.00,769.17)(28.841,9.000){2}{\rule{0.761pt}{0.400pt}}
\multiput(1272.00,779.59)(1.776,0.489){15}{\rule{1.478pt}{0.118pt}}
\multiput(1272.00,778.17)(27.933,9.000){2}{\rule{0.739pt}{0.400pt}}
\multiput(1303.00,788.59)(1.834,0.489){15}{\rule{1.522pt}{0.118pt}}
\multiput(1303.00,787.17)(28.841,9.000){2}{\rule{0.761pt}{0.400pt}}
\multiput(1335.00,797.59)(2.013,0.488){13}{\rule{1.650pt}{0.117pt}}
\multiput(1335.00,796.17)(27.575,8.000){2}{\rule{0.825pt}{0.400pt}}
\multiput(1366.00,805.59)(2.079,0.488){13}{\rule{1.700pt}{0.117pt}}
\multiput(1366.00,804.17)(28.472,8.000){2}{\rule{0.850pt}{0.400pt}}
\multiput(1398.00,813.59)(2.013,0.488){13}{\rule{1.650pt}{0.117pt}}
\multiput(1398.00,812.17)(27.575,8.000){2}{\rule{0.825pt}{0.400pt}}
\end{picture}

\smallskip

{\large \bf Fig.~3a}

\bigskip

% GNUPLOT: LaTeX picture
\setlength{\unitlength}{0.240900pt}
\ifx\plotpoint\undefined\newsavebox{\plotpoint}\fi
\begin{picture}(1500,900)(0,0)
\font\gnuplot=cmr10 at 10pt
\gnuplot
\sbox{\plotpoint}{\rule[-0.200pt]{0.400pt}{0.400pt}}%
\put(221.0,163.0){\rule[-0.200pt]{4.818pt}{0.400pt}}
\put(201,163){\makebox(0,0)[r]{-8}}
\put(1460.0,163.0){\rule[-0.200pt]{4.818pt}{0.400pt}}
\put(221.0,262.0){\rule[-0.200pt]{4.818pt}{0.400pt}}
\put(201,262){\makebox(0,0)[r]{-7}}
\put(1460.0,262.0){\rule[-0.200pt]{4.818pt}{0.400pt}}
\put(221.0,362.0){\rule[-0.200pt]{4.818pt}{0.400pt}}
\put(201,362){\makebox(0,0)[r]{-6}}
\put(1460.0,362.0){\rule[-0.200pt]{4.818pt}{0.400pt}}
\put(221.0,461.0){\rule[-0.200pt]{4.818pt}{0.400pt}}
\put(201,461){\makebox(0,0)[r]{-5}}
\put(1460.0,461.0){\rule[-0.200pt]{4.818pt}{0.400pt}}
\put(221.0,561.0){\rule[-0.200pt]{4.818pt}{0.400pt}}
\put(201,561){\makebox(0,0)[r]{-4}}
\put(1460.0,561.0){\rule[-0.200pt]{4.818pt}{0.400pt}}
\put(221.0,660.0){\rule[-0.200pt]{4.818pt}{0.400pt}}
\put(201,660){\makebox(0,0)[r]{-3}}
\put(1460.0,660.0){\rule[-0.200pt]{4.818pt}{0.400pt}}
\put(221.0,760.0){\rule[-0.200pt]{4.818pt}{0.400pt}}
\put(201,760){\makebox(0,0)[r]{-2}}
\put(1460.0,760.0){\rule[-0.200pt]{4.818pt}{0.400pt}}
\put(221.0,859.0){\rule[-0.200pt]{4.818pt}{0.400pt}}
\put(201,859){\makebox(0,0)[r]{-1}}
\put(1460.0,859.0){\rule[-0.200pt]{4.818pt}{0.400pt}}
\put(221.0,163.0){\rule[-0.200pt]{0.400pt}{4.818pt}}
\put(221,122){\makebox(0,0){20}}
\put(221.0,839.0){\rule[-0.200pt]{0.400pt}{4.818pt}}
\put(378.0,163.0){\rule[-0.200pt]{0.400pt}{4.818pt}}
\put(378,122){\makebox(0,0){40}}
\put(378.0,839.0){\rule[-0.200pt]{0.400pt}{4.818pt}}
\put(536.0,163.0){\rule[-0.200pt]{0.400pt}{4.818pt}}
\put(536,122){\makebox(0,0){60}}
\put(536.0,839.0){\rule[-0.200pt]{0.400pt}{4.818pt}}
\put(693.0,163.0){\rule[-0.200pt]{0.400pt}{4.818pt}}
\put(693,122){\makebox(0,0){80}}
\put(693.0,839.0){\rule[-0.200pt]{0.400pt}{4.818pt}}
\put(851.0,163.0){\rule[-0.200pt]{0.400pt}{4.818pt}}
\put(851,122){\makebox(0,0){100}}
\put(851.0,839.0){\rule[-0.200pt]{0.400pt}{4.818pt}}
\put(1008.0,163.0){\rule[-0.200pt]{0.400pt}{4.818pt}}
\put(1008,122){\makebox(0,0){120}}
\put(1008.0,839.0){\rule[-0.200pt]{0.400pt}{4.818pt}}
\put(1165.0,163.0){\rule[-0.200pt]{0.400pt}{4.818pt}}
\put(1165,122){\makebox(0,0){140}}
\put(1165.0,839.0){\rule[-0.200pt]{0.400pt}{4.818pt}}
\put(1323.0,163.0){\rule[-0.200pt]{0.400pt}{4.818pt}}
\put(1323,122){\makebox(0,0){160}}
\put(1323.0,839.0){\rule[-0.200pt]{0.400pt}{4.818pt}}
\put(1480.0,163.0){\rule[-0.200pt]{0.400pt}{4.818pt}}
\put(1480,122){\makebox(0,0){180}}
\put(1480.0,839.0){\rule[-0.200pt]{0.400pt}{4.818pt}}
\put(221.0,163.0){\rule[-0.200pt]{303.293pt}{0.400pt}}
\put(1480.0,163.0){\rule[-0.200pt]{0.400pt}{167.666pt}}
\put(221.0,859.0){\rule[-0.200pt]{303.293pt}{0.400pt}}
\put(1,511){\makebox(0,0){$\Delta \psi^2_1 \times 10^2$}}
\put(850,21){\makebox(0,0){$\mu_f$ (GeV)}}
\put(221.0,163.0){\rule[-0.200pt]{0.400pt}{167.666pt}}
\put(265,767){\usebox{\plotpoint}}
\multiput(265.58,764.39)(0.497,-0.662){59}{\rule{0.120pt}{0.629pt}}
\multiput(264.17,765.69)(31.000,-39.694){2}{\rule{0.400pt}{0.315pt}}
\multiput(296.58,723.72)(0.497,-0.562){61}{\rule{0.120pt}{0.550pt}}
\multiput(295.17,724.86)(32.000,-34.858){2}{\rule{0.400pt}{0.275pt}}
\multiput(328.58,687.87)(0.497,-0.515){59}{\rule{0.120pt}{0.513pt}}
\multiput(327.17,688.94)(31.000,-30.935){2}{\rule{0.400pt}{0.256pt}}
\multiput(359.00,656.92)(0.571,-0.497){53}{\rule{0.557pt}{0.120pt}}
\multiput(359.00,657.17)(30.844,-28.000){2}{\rule{0.279pt}{0.400pt}}
\multiput(391.00,628.92)(0.596,-0.497){49}{\rule{0.577pt}{0.120pt}}
\multiput(391.00,629.17)(29.803,-26.000){2}{\rule{0.288pt}{0.400pt}}
\multiput(422.00,602.92)(0.668,-0.496){45}{\rule{0.633pt}{0.120pt}}
\multiput(422.00,603.17)(30.685,-24.000){2}{\rule{0.317pt}{0.400pt}}
\multiput(454.00,578.92)(0.706,-0.496){41}{\rule{0.664pt}{0.120pt}}
\multiput(454.00,579.17)(29.623,-22.000){2}{\rule{0.332pt}{0.400pt}}
\multiput(485.00,556.92)(0.740,-0.496){39}{\rule{0.690pt}{0.119pt}}
\multiput(485.00,557.17)(29.567,-21.000){2}{\rule{0.345pt}{0.400pt}}
\multiput(516.00,535.92)(0.847,-0.495){35}{\rule{0.774pt}{0.119pt}}
\multiput(516.00,536.17)(30.394,-19.000){2}{\rule{0.387pt}{0.400pt}}
\multiput(548.00,516.92)(0.866,-0.495){33}{\rule{0.789pt}{0.119pt}}
\multiput(548.00,517.17)(29.363,-18.000){2}{\rule{0.394pt}{0.400pt}}
\multiput(579.00,498.92)(1.009,-0.494){29}{\rule{0.900pt}{0.119pt}}
\multiput(579.00,499.17)(30.132,-16.000){2}{\rule{0.450pt}{0.400pt}}
\multiput(611.00,482.92)(0.977,-0.494){29}{\rule{0.875pt}{0.119pt}}
\multiput(611.00,483.17)(29.184,-16.000){2}{\rule{0.438pt}{0.400pt}}
\multiput(642.00,466.92)(1.079,-0.494){27}{\rule{0.953pt}{0.119pt}}
\multiput(642.00,467.17)(30.021,-15.000){2}{\rule{0.477pt}{0.400pt}}
\multiput(674.00,451.92)(1.121,-0.494){25}{\rule{0.986pt}{0.119pt}}
\multiput(674.00,452.17)(28.954,-14.000){2}{\rule{0.493pt}{0.400pt}}
\multiput(705.00,437.92)(1.158,-0.494){25}{\rule{1.014pt}{0.119pt}}
\multiput(705.00,438.17)(29.895,-14.000){2}{\rule{0.507pt}{0.400pt}}
\multiput(737.00,423.92)(1.210,-0.493){23}{\rule{1.054pt}{0.119pt}}
\multiput(737.00,424.17)(28.813,-13.000){2}{\rule{0.527pt}{0.400pt}}
\multiput(768.00,410.92)(1.358,-0.492){21}{\rule{1.167pt}{0.119pt}}
\multiput(768.00,411.17)(29.579,-12.000){2}{\rule{0.583pt}{0.400pt}}
\multiput(800.00,398.92)(1.315,-0.492){21}{\rule{1.133pt}{0.119pt}}
\multiput(800.00,399.17)(28.648,-12.000){2}{\rule{0.567pt}{0.400pt}}
\multiput(831.00,386.92)(1.486,-0.492){19}{\rule{1.264pt}{0.118pt}}
\multiput(831.00,387.17)(29.377,-11.000){2}{\rule{0.632pt}{0.400pt}}
\multiput(863.00,375.92)(1.439,-0.492){19}{\rule{1.227pt}{0.118pt}}
\multiput(863.00,376.17)(28.453,-11.000){2}{\rule{0.614pt}{0.400pt}}
\multiput(894.00,364.92)(1.486,-0.492){19}{\rule{1.264pt}{0.118pt}}
\multiput(894.00,365.17)(29.377,-11.000){2}{\rule{0.632pt}{0.400pt}}
\multiput(926.00,353.92)(1.590,-0.491){17}{\rule{1.340pt}{0.118pt}}
\multiput(926.00,354.17)(28.219,-10.000){2}{\rule{0.670pt}{0.400pt}}
\multiput(957.00,343.92)(1.642,-0.491){17}{\rule{1.380pt}{0.118pt}}
\multiput(957.00,344.17)(29.136,-10.000){2}{\rule{0.690pt}{0.400pt}}
\multiput(989.00,333.93)(1.776,-0.489){15}{\rule{1.478pt}{0.118pt}}
\multiput(989.00,334.17)(27.933,-9.000){2}{\rule{0.739pt}{0.400pt}}
\multiput(1020.00,324.93)(1.834,-0.489){15}{\rule{1.522pt}{0.118pt}}
\multiput(1020.00,325.17)(28.841,-9.000){2}{\rule{0.761pt}{0.400pt}}
\multiput(1052.00,315.93)(1.776,-0.489){15}{\rule{1.478pt}{0.118pt}}
\multiput(1052.00,316.17)(27.933,-9.000){2}{\rule{0.739pt}{0.400pt}}
\multiput(1083.00,306.93)(1.776,-0.489){15}{\rule{1.478pt}{0.118pt}}
\multiput(1083.00,307.17)(27.933,-9.000){2}{\rule{0.739pt}{0.400pt}}
\multiput(1114.00,297.93)(2.079,-0.488){13}{\rule{1.700pt}{0.117pt}}
\multiput(1114.00,298.17)(28.472,-8.000){2}{\rule{0.850pt}{0.400pt}}
\multiput(1146.00,289.93)(2.013,-0.488){13}{\rule{1.650pt}{0.117pt}}
\multiput(1146.00,290.17)(27.575,-8.000){2}{\rule{0.825pt}{0.400pt}}
\multiput(1177.00,281.93)(2.079,-0.488){13}{\rule{1.700pt}{0.117pt}}
\multiput(1177.00,282.17)(28.472,-8.000){2}{\rule{0.850pt}{0.400pt}}
\multiput(1209.00,273.93)(2.013,-0.488){13}{\rule{1.650pt}{0.117pt}}
\multiput(1209.00,274.17)(27.575,-8.000){2}{\rule{0.825pt}{0.400pt}}
\multiput(1240.00,265.93)(2.399,-0.485){11}{\rule{1.929pt}{0.117pt}}
\multiput(1240.00,266.17)(27.997,-7.000){2}{\rule{0.964pt}{0.400pt}}
\multiput(1272.00,258.93)(2.013,-0.488){13}{\rule{1.650pt}{0.117pt}}
\multiput(1272.00,259.17)(27.575,-8.000){2}{\rule{0.825pt}{0.400pt}}
\multiput(1303.00,250.93)(2.399,-0.485){11}{\rule{1.929pt}{0.117pt}}
\multiput(1303.00,251.17)(27.997,-7.000){2}{\rule{0.964pt}{0.400pt}}
\multiput(1335.00,243.93)(2.323,-0.485){11}{\rule{1.871pt}{0.117pt}}
\multiput(1335.00,244.17)(27.116,-7.000){2}{\rule{0.936pt}{0.400pt}}
\multiput(1366.00,236.93)(2.399,-0.485){11}{\rule{1.929pt}{0.117pt}}
\multiput(1366.00,237.17)(27.997,-7.000){2}{\rule{0.964pt}{0.400pt}}
\multiput(1398.00,229.93)(2.751,-0.482){9}{\rule{2.167pt}{0.116pt}}
\multiput(1398.00,230.17)(26.503,-6.000){2}{\rule{1.083pt}{0.400pt}}
\end{picture}

\smallskip

{\large \bf Fig.~3b}

\end{center}

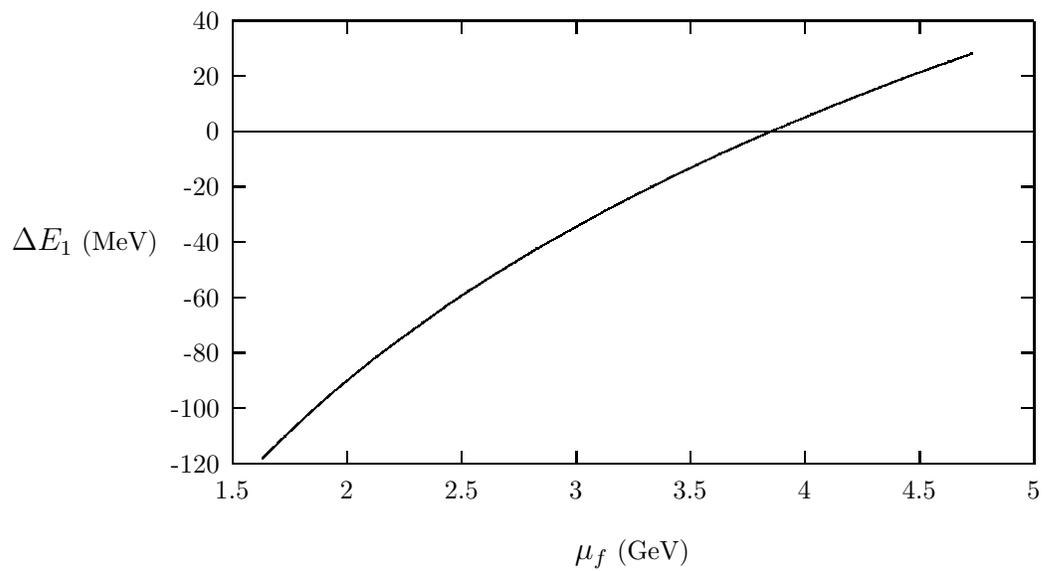
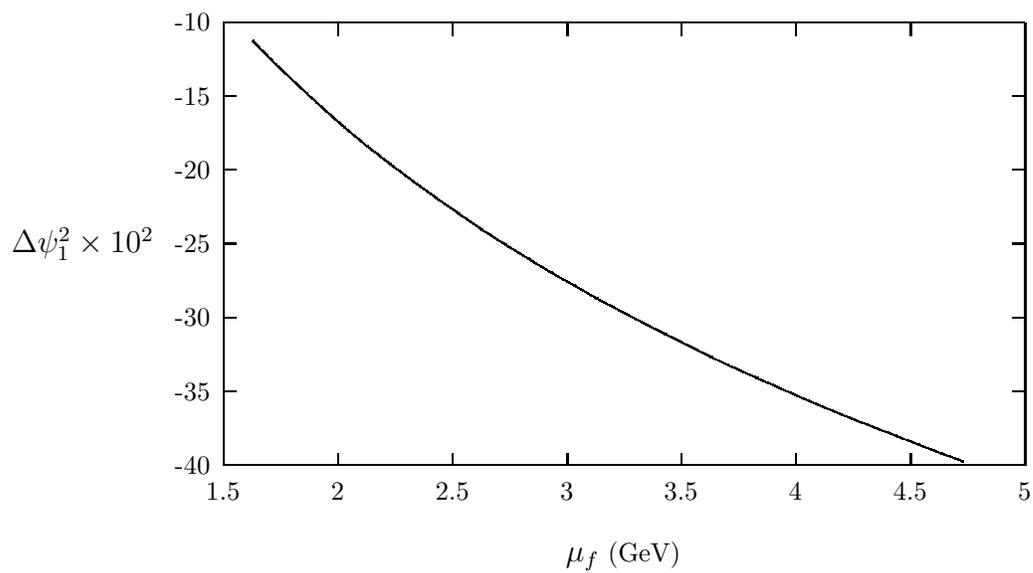
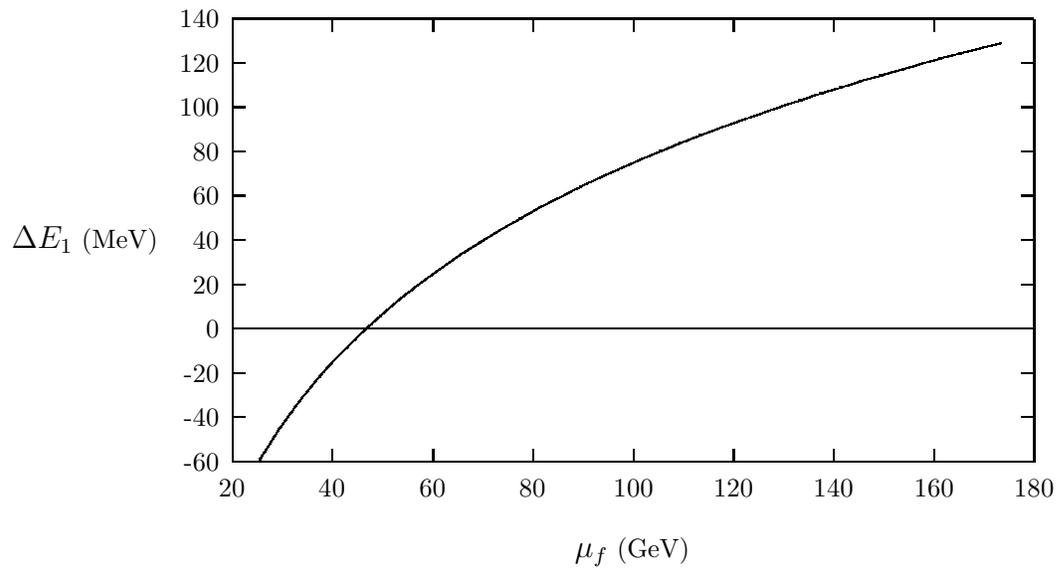
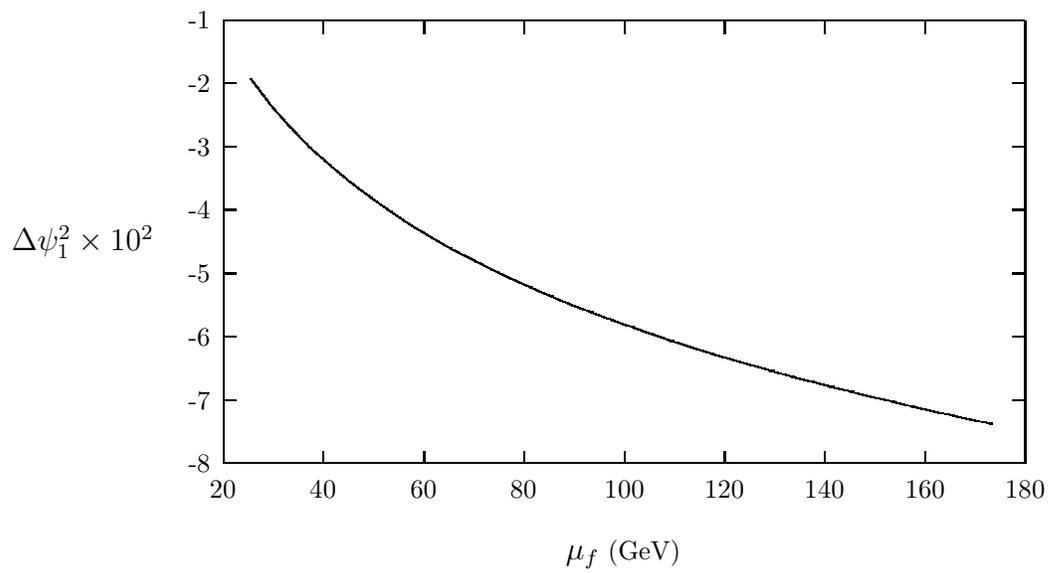
\end{document}